\documentclass[11pt,a4paper]{article}

\pdfoutput=1
\usepackage{jheppub}
\usepackage{graphicx}
\usepackage{amssymb}
\usepackage{amsmath}
\usepackage{mathtools}
\usepackage{bm}
\usepackage{latexsym}
\usepackage{epsfig}
\usepackage{float}
\usepackage{textcomp}
\usepackage{color}
\usepackage{float}
\usepackage{blindtext}
\usepackage{enumitem}
\usepackage{xcolor}
\usepackage[section]{placeins}
\usepackage{float}
\usepackage{tikz} % To generate the plot from csv
\usepackage{pgfplots}
\usepackage{graphicx}
\pgfplotsset{compat=newest} % Allows to place the legend below plot
\usepgfplotslibrary{units} % Allows to enter the units nicely

\usepackage{calligra}
\DeclareMathAlphabet{\mathcalligra}{T1}{calligra}{m}{n}
\DeclareFontShape{T1}{calligra}{m}{n}{<->s*[2.2]callig15}{}

\usepackage[mathscr]{euscript}
\usepackage[export]{adjustbox}
\usepackage{amsmath}

\makeatletter
\gdef\@fpheader{}
\makeatother

\def\f{\frac}

\def\d{{\rm d}}
\def\Mpl{M_{_{\rm Pl}}}
\def\beq{\begin{equation}}
\def\eeq{\end{equation}} 
\def\be{\begin{eqnarray}}
\def\ee{\end{eqnarray}}

\def\MPBH{M_{\rm PBH}}

\definecolor{lime}{HTML}{A6CE39}
\DeclareRobustCommand{\orcidicon}{
	\begin{tikzpicture}
	\draw[lime, fill=lime] (0,0) 
	circle [radius=0.2] 
	node[white] {{\fontfamily{qag}\selectfont \tiny ID}};
	\draw[white, fill=white] (-0.0625,0.095) 
	circle [radius=0.007];
	\end{tikzpicture}
	\hspace{-2mm}
}

\def\be{\begin{equation}}
\def\ee{\end{equation}} 
\def\bea{\begin{eqnarray}}
\def\eea{\end{eqnarray}}

\def\Mpl{M_{P}}
\def\f{\frac}

\def\d{\mathrm{d}}

\def\lsim{\mathrel{\rlap{\lower4pt\hbox{\hskip0.5pt$\sim$}}
 \raise1pt\hbox{$<$}}}         %less than or approx. symbol
\def\gsim{\mathrel{\rlap{\lower4pt\hbox{\hskip0.5pt$\sim$}}
 \raise1pt\hbox{$>$}}}         %greater than or approx. symbol

\newcommand{\dd}{\mathrm{d}}

\newcommand{\ba}{\begin{aligned}}
\newcommand{\ea}{\end{aligned}}

\newcommand{\x}{\mathbf{x}}

\def\Min{M_{\mathrm {\mathrm PBH}}}
\def\Mpbh{M_{\mathrm {\mathrm PBH}}}

\def\f{\frac}

\def\d{{\rm d}}
\def\Mpl{M_{_{\rm Pl}}}
\def\be{\begin{equation}}
\def\ee{\end{equation}} 
\def\bea{\begin{eqnarray}}
\def\eea{\end{eqnarray}}

\foreach \x in {A, ..., Z}{\expandafter\xdef\csname orcid\x\endcsname{\noexpand\href{https://orcid.org/\csname orcidauthor\x\endcsname}
			{\noexpand\orcidicon}}
}

\def\f{\frac}

\def\d{{\rm d}}
\def\Mpl{M_{\rm Pl}}
\def\be{\begin{equation}}
\def\ee{\end{equation}} 
\def\bea{\begin{eqnarray}}
\def\eea{\end{eqnarray}}

\numberwithin{equation}{section}

\begin{document}

\title{Memory burden effect mimics reheating signatures on SGWB from ultra-low mass PBH domination}
\author[a]{Nilanjandev Bhaumik\orcidA{}}
\emailAdd{nilanjandevbhaumik@gmail.com}
\affiliation[a]{International Centre for Theoretical Physics Asia-Pacific, Beijing 100190, China}
\author[b, c]{, Md Riajul Haque\orcidB{}}
\emailAdd{riaj.0009@gmail.com}
\affiliation[b]{Centre for Strings, Gravitation, and Cosmology,
Department of Physics, Indian Institute of Technology Madras, 
Chennai~600036, India}
\affiliation[c]{Physics and Applied Mathematics Unit, Indian Statistical Institute, 203 B.T. Road, Kolkata 700108, India}
\author[d]{, Rajeev Kumar Jain\orcidC{}}
 \emailAdd{rkjain@iisc.ac.in}
\affiliation[d]{
Department of Physics, Indian Institute of Science,
C. V. Raman Road, Bangalore 560012, India}
\author[e]{\\and Marek Lewicki\orcidD{}}
\emailAdd{marek.lewicki@fuw.edu.pl}
\affiliation[e]{
Faculty of Physics, University of Warsaw, ul. Pasteura 5, 02-093 Warsaw, Poland
}%

\date{\today}

\abstract{
Ultra-low mass primordial black holes (PBH), briefly dominating the expansion of the universe, would leave detectable imprints in the secondary stochastic gravitational wave background (SGWB). Such a scenario leads to a characteristic doubly peaked spectrum of SGWB and strongly depends on the Hawking evaporation of such light PBHs. However, these observable signatures are significantly altered if the memory burden effect during the evaporation of PBHs is taken into account. 
We show that for the SGWB induced by PBH density fluctuations, the memory burden effects on the Hawking evaporation of ultra-low mass PBHs can mimic the signal arising due to the non-standard reheating epoch before PBH domination.
Finally, we point out that this degeneracy can be broken by the simultaneous detection of the first peak in the SGWB, which is typically induced by the inflationary adiabatic perturbations.
}
\keywords{Primordial black hole, Reheating, Memory burden effect, Stochastic gravitational wave background}
\maketitle

%\tableofcontents

%%######################################################################################%%

\section{Introduction}

Primordial black holes (PBH)~\cite{Zeldovich:1967lct,Hawking:1971ei,Carr:1974nx, Khlopov:1985fch} which may have been produced in the very early universe due to the collapse of overdense primordial fluctuations, can play a crucial role in its subsequent evolution. Over the years, PBHs have gained enormous attention as an important tool in understanding and probing the early universe cosmology and particle physics~\cite{Escriva:2022duf,LISACosmologyWorkingGroup:2023njw}. PBHs are also one of the most popular dark matter (DM) candidates and can constitute the entire DM in the asteroid mass range~\cite{Carr:2021bzv}. Moreover, Hawking evaporation~\cite{Hawking:1974rv} from PBHs can be a very important probe for quantum gravity theories \cite{Rovelli:1997yv,Mathur:2011wg,Papanikolaou:2021uhe,Calmet:2023gbw}.

The Hawking evaporation of (primordial) black holes is usually described by the semiclassical formalism  \cite{Hawking:1974rv,Page:1976dff}, which suggests that a black hole radiates a thermal spectrum of particles, with temperature  $T \sim 1/M_{\rm PBH}$. 
Moreover, the semiclassical description is self-similar, and the evaporation process terminates when $M_{\rm PBH} \to 0$. This semiclassical formalism leads to a constraint on $M_{\rm PBH} \sim 10^{15}\, {\rm g}$, which indicates that PBHs lighter than this mass would be entirely evaporated by the present epoch and would not constitute any fraction of the DM today. Similarly, ultra-low mass PBHs with masses below $\sim 10^{9}\, {\rm g}$ would evaporate due to their Hawking radiation far before the Big Bang nucleosynthesis (BBN), and can briefly dominate the universe as there are no effective constraints on the abundance of these ultra-low mass PBHs ~\cite{Carr:2020gox,Auffinger:2022khh} (see also a recent work \cite{Boccia:2024nly} suggesting a smaller PBH mass bound). Such a scenario would not only affect the thermal history of the universe but also lead to very distinct observable imprints, such as a resonant or poltergeist amplification \cite{Inomata:2019zqy, Inomata:2019ivs} of the stochastic gravitational wave background (SGWB) comprised of resonant SGWB peaks due to inflationary adiabatic perturbations~\cite{Inomata:2020lmk, White:2021hwi, Bhaumik:2022pil, Bhaumik:2022zdd, Bhaumik:2023wmw} and the PBH density fluctuations~\cite{Papanikolaou:2020qtd, Domenech:2020ssp, Dalianis:2020gup, Domenech:2021ztg, Domenech:2021wkk, Papanikolaou:2021uhe, Bhaumik:2022pil, Bhaumik:2022zdd,Papanikolaou:2022chm}. In this scenario, the standard radiation domination (RD) is achieved by the decay of the PBHs \cite{Carr:1976zz,Hidalgo:2011fj,Martin:2019nuw, Hooper:2019gtx, Hooper:2020evu, Hooper:2020otu, Bernal:2020bjf, Cheek:2021cfe, Cheek:2022mmy, Mazde:2022sdx,RiajulHaque:2023cqe, Calabrese:2023key, Barman:2024slw}.

Recently, it has been pointed out that the semiclassical description of Hawking evaporation can not hold over the entire black hole lifetime, and one must take into account the effects of quantum backreaction on the black hole over the evaporation timescales. An efficient way to take this backreaction into account is through the so-called "memory burden" effect, which assumes a black hole as a saturated coherent state or a Bose–Einstein condensate bound state of soft gravitons \cite{Dvali:2011aa,Dvali:2012en,Dvali:2015rea, Dvali:2024hsb}. This suggests that the semiclassical formalism of Hawking evaporation breaks down at the latest by the time a black hole loses half of its mass due to Hawking evaporation \cite{Dvali:2020wft,Dvali:2018xpy}. The microscopic energy states, which were gapless at the start of Hawking evaporation, allowing a maximal storage capacity of information, can no longer stay gapless after the mass of the black hole reduces significantly from its initial value. Thus, the dynamics of Hawking evaporation of a black hole after losing half of its initial mass $\Mpbh$ is in no way identical to another newly formed black hole with initial mass $\Mpbh/2$ \cite{Dvali:2020wft}. Such a crucial deviation of the Hawking evaporation formalism from the semiclassical approximation proves to have far-reaching implications for the candidature of PBH as DM~\cite{Alexandre:2024nuo, Thoss:2024hsr,Haque:2024eyh,Dvali:2024hsb,Tamta:2024pow} and ultra-low mass PBH domination scenario~\cite{Balaji:2024hpu}.

A recent study of the memory burden effect on the ultra-low mass PBH domination scenario shows interesting signatures in the resulting SGWB peaks \cite{Balaji:2024hpu} originating from the PBH density fluctuations, which are initially isocurvature perturbations that turn to adiabatic perturbations at the time of PBH domination. These adiabatic perturbations then contribute to the second-order resonant (or poltergeist) SGWB \cite{Inomata:2019ivs, Inomata:2019zqy} (also see \cite{Pearce:2023kxp})  after PBHs evaporate and the standard RD starts. In this work, we also study this particular contribution and find that for the PBH density fluctuation-induced SGWB peak, the modification induced by the memory burden effect can be mimicked by considering different possible reheating histories that can precede the PBH domination. 

Our existing observational probes provide us with no definitive clue to understanding the expansion history or constituents of our universe between the end of inflation and BBN~\cite{Allahverdi:2020bys}. Thus it is possible to consider a non-standard reheating phase 
between the end of inflation and PBH domination. We assume PBHs form during this reheating phase and eventually dominate but evaporate before BBN. The variation in the reheating history leaves a strong impact on the SGWB spectra, and we find the effect of considering non-standard reheating history and memory burden effect to have indistinguishable SGWB signatures for PBH density fluctuation induced SGWB. This property of one scenario to mimic the other makes it impossible to identify these scenarios from possible future SGWB detections. Thus only this peak is not sufficient to identify the memory burden parameters. On the other hand, if we also consider the inflationary adiabatic perturbation-induced peak for both of these scenarios, this degeneracy can be broken. We further discuss other possible ways to break this degeneracy.

This paper is organized as follows. In section \ref{sec-reh}, we study the implications on the SGWB with a non-standard reheating history before the epoch of PBH domination. In section \ref{sec-mbe}, we consider the memory burden effect in the context of ultra-low mass PBH domination. 
Section~\ref{sec-deg} focuses on the degeneracy between these two effects and what are the available means to break the degeneracies.
In section \ref{sec-cd}, we summarise our results and discuss future possibilities. We denote the time-dependent PBH mass as $M$ while the initial PBH mass is expressed as $M_{\rm PBH}$. The reduced Planck mass is defined as $\Mpl^2 = 1/8\pi G$.

\section{SGWB for a non-standard reheating history before PBH domination}
\label{sec-reh}

After the end of the inflation, the inflaton field oscillates around the minima of the potential. Depending on the nature of the potential at the minima, it can generate an equation of state (EOS) $w$ different from radiation ($w=1/3$) and depending on the rate of decay of the inflaton to standard model particles, this reheating can last for quite a prolonged duration. In the standard reheating scenario, the RD era is achieved through the decay of the oscillating inflaton field, whose energy dominates the total energy budget of the universe before it decays. However, in our current scenario, we assume a different possibility of the background evolution wherein PBHs are formed during this inflaton-dominated universe. After the formation, these PBHs follow the expansion rate of the nonrelativistic matter. Thus, the energy density dilutes slower than the total energy density when we consider the effective EOS, $w$, during the inflaton oscillation and decay to be greater than zero. As a result, PBHs may dominate the universe before they decay if their initial abundance is large enough. Therefore, we consider a PBH domination epoch sandwiched between an inflaton decay period with EOS $w$ and standard  RD.  For phenomenological purposes, without going into the details of the inflaton decay dynamics, we assume a $w$-dominated phase after inflation. Note that, to achieve such a scenario, the inflaton lifetime must be greater than the time scale to reach PBH domination. While the inflaton decay phase can be called reheating, the PBH evaporation, which ultimately leads to the standard model particles, can also be termed reheating. To avoid ambiguity in this work, we call the inflaton decay phase non-standard reheating or simply reheating, while the PBH evaporation event would be referred to as the start of RD.

We assume a population of ultra-low mass non-spinning PBHs with monochromatic mass range to form at $\tau=\tau_f$ during a non-standard reheating phase, dominate the universe at $\tau=\tau_m$ and eventually decay due to Hawking evaporation at $\tau=\tau_r$. For such a reheating phase, using the Friedmann equations and continuity equation, the scale factor can be expressed as,
\begin{align}
    a(\tau)=\left( c_0 \tau + c_1\right)^{\frac{2}{3 w+1}}, \hspace{0.5cm}  a(t)=\left( c_2 t + c_3\right)^{\frac{2}{3 (w+1)}} \, .
\end{align}
The existence of a non-standard reheating history alters the rate of expansion of the universe between PBH formation and domination. Thus, the background dynamics are different from the usual scenario wherein the phase between the end of inflation and the PBH domination is assumed to be radiation-dominated, which is usually referred to as early radiation domination (eRD). The comoving wavenumbers that re-enter the horizon during PBH evaporation ($k_r$), PBH domination ($k_m$), and PBH formation ($k_f$) in the case of eRD
have been derived in our earlier work~\cite{Bhaumik:2022pil} in terms of PBH mass $\MPBH$ and the initial abundance $\beta_f$~\cite{Bhaumik:2022pil}. 
%We denote the comoving wavenumber values in this special case with a subscript "0" and 
However, in the current scenario, there arise modifications due to the non-standard reheating phase with EOS $w$, which can be expressed as,
\begin{eqnarray}
k_{r}&\approx&\left(\frac{2.1 \times 10^{11}}{ \text{Mpc}}\right) \left( \frac{M_{\rm PBH}}{10^4\, {\rm g} } \right)^{-\frac{3}{2}},\\
k_{m}&\approx&\left(\frac{ 8.7 \times 10^{17} }{ \text{Mpc}}\right)\left( \frac{M_{\rm PBH}}{10^4\, {\rm g} } \right)^{-\frac{5}{6}} \beta_f^{\frac{1 + w}{6 w}} w^{\frac{1}{2}}\left(w+\frac{1}{3}\right),\hspace{.7cm}\\
k_{f}&\approx&\left(\frac{ 8.7 \times 10^{17}  }{ \text{Mpc}}\right)\left( \frac{M_{\rm PBH}}{10^4\, {\rm g} } \right)^{-\frac{5}{6}} \beta_f^{-\frac{1}{3}} w^{\frac{1}{2}}\left(w+\frac{1}{3}\right), \hspace{.7cm}
 \label{waves}
\end{eqnarray}
and the duration of PBH domination is quantified with $\tau_{\rm rat} \equiv \tau_r /\tau_m$. Note that both $k_m$ and $k_f$ depend on $w$, but $k_r$ is independent of $w$.

In the context of ultra-low mass PBH domination and evaporation, we consider the generation of second-order SGWB just after PBH evaporation through the resonant amplification or the “poltergeist mechanism”~\cite{Inomata:2019ivs}. At the onset of standard RD, two different sources of adiabatic perturbations become important~\cite{Bhaumik:2022pil}. One is the inflationary adiabatic contribution, which we model with the standard nearly scale-invariant scalar power spectra, 
\begin{align}
\label{power-law}
\mathcal{P}_{\mathcal{R}}=A_s \left( \frac{k}{k_{p}} \right)^{n_s -1}.
\end{align}
with amplitude $A_s \simeq 2.09 \times 10^{-9} $ and scalar index $n_s \simeq 0.965$  consistent with Planck 2018 at pivot scale $k_p=0.05$  ${\rm Mpc}^{-1}$ \cite{Planck:2018jri}.
The other contribution comes from the isocurvature fluctuations of PBH density perturbations, which become adiabatic perturbations by the time PBHs dominate~\cite{Papanikolaou:2020qtd, Domenech:2021wkk}. During PBH domination, these isocurvature-induced adiabatic fluctuations dominate over the inflationary adiabatic perturbations on small scales, but we assume the large-scale scalar perturbations to follow the nearly scale-invariant inflationary perturbations profile consistent with CMB observations at large scales. Assuming a Poissonian distribution of PBHs, the matter density perturbation power spectrum at the time of PBH formation can be expressed as \cite{Papanikolaou:2020qtd, Domenech:2021wkk},
\begin{align}
    \mathcal{P}_{\delta_{m}}(k,\tau_f)= \frac{2}{3\pi} \left(\frac{k}{k_{\rm UV}}\right)^3 \, ,
\end{align}
where the cutoff scale of PBH density perturbations $k_{\rm UV}$ can be expressed as,
\begin{eqnarray}
\label{Eq:Kuv}
k_{\rm UV}&=k_f \beta_f^{1/3} w^{-1/2}& \nonumber \\ &\approx 8.7 \times 10^{17} &\left( \frac{M_{\rm PBH}}{10^4 {\rm g} } \right)^{-\frac{5}{6}}\left(w+\frac{1}{3}\right)\, \text{Mpc}^{-1}  \, ,
\end{eqnarray}
which refers to the average moving mean distance between two PBHs. For modes with $k \ge k_{\rm UV}$, the granularity of PBHs becomes important. Superhorizon modes of the PBH density contrast can be related to Bardeen potential at the time of PBH evaporation easily,
\begin{equation}
\Phi(k, \tau_r)\vert_{\rm PBH}=\frac{1}{5} \delta_m(k, \tau_f)\equiv \frac{1}{5} \delta_i(k) \, ,
\end{equation}
where we assume the matter density perturbation due to PBHs at the time of PBH formation to be $\delta_i(k)$. On the other hand, the subhorizon modes of the density contrast go through the time evolution, given by a general-EOS version of Mészáros equation  \cite{Meszaros:1974tb,Mukhanov:2005sc},
\begin{equation}
s^2 (1+s^{-3w}) \frac{d^2\delta_m(k,s)}{ds^2}+\frac{3}{2}s(1+(1-w)s^{-3w}) \frac{d\delta_m(k,s)}{ds}-\frac{3}{2}\delta_m(k,s)=0,
\label{isoeq}
\end{equation}
where $s\equiv a/a_m$.  Solving Eq.~\eqref{isoeq} we get, 
\begin{eqnarray}
\delta_m(k,s)= s^{\frac{-1 + 3 w}{2}} {}_{2}F_1\bigg[\frac{1}{2} - \frac{1}{2 w}, \frac{1}{2} + \frac{1}{3 w} , 
 \frac{3}{2} - \frac{1}{6 w}, -s^{3 w}\bigg] C_2 \nonumber \\+ \; {}_{2}F_{1}\bigg[-\frac{1}{3 w}, \frac{1}{2 w}, \frac{1}{2} + \frac{1}{6 w}, -s^{3 w}\bigg] C_1 
\hspace{1cm}\end{eqnarray}
The modes we consider are superhorizon modes at the time of PBH formation $s=s_f \equiv a_f/a_m$. Thus we can assume $\delta_m(k,s_f)=\delta_i(k)$ and $ \frac{d\delta_m(k,s)}{ds} \vert{s=s_f} \approx 0$, where $s_f \ll 1$ . This allows us to neglect $C_2$ as $C_2 \ll C_1$ and at the time of PBH domination at $s=1$ we can get ,
\begin{align}
\delta_m (k,s=1)={}_{2}F_{1}\bigg[-\frac{1}{3 w}, \frac{1}{2 w}, \frac{1}{2} + \frac{1}{6 w}, -1\bigg]\delta_i(k) \, .
\end{align}
Thus at the time of PBH formation for subhorizon modes with comoving wavenumber $k > k_m$, we can express  $\Phi(k)\vert_{\rm PBH}$ with  
\begin{equation}
\Phi(k,\tau_m)\vert_{\rm PBH}=\frac{3}{2} \left(\frac{k_m}{k}\right)^2 {}_{2}F_{1}\bigg[-\frac{1}{3 w}, \frac{1}{2 w}, \frac{1}{2} + \frac{1}{6 w}, -1\bigg]\delta_i (k)\, .
\end{equation}

Now we have the two components of adiabatic perturbations at the time of PBH domination,  the inflationary adiabatic component and the PBH density fluctuation-driven isocurvature turned adiabatic component,
\begin{equation}
\Phi(k,\tau_m)= \Phi_k(k,\tau_m)\vert_{\rm infl}+ \Phi(k,\tau_m)\vert_{\rm PBH} \, .
\end{equation}
It is necessary to understand the time evolution of the first-order scalar perturbation $\Phi(k, \tau_m)$ during the PBH domination and its final value at the end of PBH domination $\Phi(k, \tau_r)$. It is this value that crucially determines the SGWB generated at the end of PBH domination or at the onset of standard RD. One interesting point to note in this context is that, though at the start of PBH domination, adiabatic scalar perturbation contains two components from two different origins, at the time of PBH domination, the isocurvature perturbation has already converted completely to the adiabatic component, and thus we can assume the system purely adiabatic. 

Now, when we consider a transition from a general$-w$ dominated universe to PBH domination, the non-dust component decays following their EOS, and the suppression factor differs from the one for radiation. In such a scenario around the PBH domination, we can imagine the universe mainly dominated by a two-component fluid: the inflaton with a constant effective EOS $w$ and the PBHs. As time passes, the PBH decay produces the radiation component. Thus, we consider three different density perturbations and velocity perturbation components and denote them in the synchronous gauge as $\delta_w$, $\delta_r$, $\delta_m$ and $\theta_w$, $\theta_r$, $\theta_m$, respectively. While for PBH, we can assume the velocity perturbation $\theta_m \approx 0$. Interestingly, as we do not consider the inflaton decay to produce the radiation particles, the background energy density of the inflaton simply dilutes due to the Hubble expansion and eventually becomes subdominant. Thus, the background energy densities follow~\cite{Poulin:2016nat,Inomata:2020lmk},
\begin{align}
\label{bck1}
  \rho'_\text{m} &= - \left( 3\mathcal H - \frac{\dd \, \text{ln}\, M_\text{PBH}}{\dd \eta} \right) \rho_\text{m}, \\
  \rho'_\text{r} &= -4\mathcal H \rho_\text{r} -\frac{\dd \, \text{ln}\, M_\text{PBH}}{\dd \eta} \rho_\text{m}, \\
  \rho'_{w} &= -3(1+w)\mathcal H \rho_{w}.
\end{align}
The equations of motion for the metric perturbations can be written as~\cite{Ma:1995ey,Inomata:2020lmk},
\begin{align} 
k^2 \epsilon - \frac{1}{2} \frac{a'}{a} \gamma' 
  &= -\frac{3}{2} \mathcal H^2 \left( \frac{\rho_\text{m}}{\rho_\text{tot} } \delta_\text{m} + \frac{\rho_\text{r}}{\rho_\text{tot}} \delta_\text{r} + \frac{\rho_\text{w}}{\rho_\text{tot}} \delta_\text{w}\right), \\
  k^2 \epsilon' 
  & = 2 \mathcal H^2 \left( \frac{\rho_\text{r}}{\rho_\text{tot}} \theta_\text{r} + \frac{\rho_\text{w}}{\rho_\text{tot}} \theta_\text{w}\right).
\end{align}
and the density and velocity perturbations follow \cite{Ma:1995ey,Poulin:2016nat,Inomata:2020lmk}:
\begin{align}
  \delta'_\text{m} &= - \frac{\gamma'}{2},\\
  \delta'_\text{r} &= -\frac{4}{3} \left(\theta_\text{r} + \gamma'/2 \right) - \frac{\dd\, \text{ln} \, M_\text{PBH}}{\dd \eta} \frac{\rho_\text{m}}{\rho_\text{r}} \left(\delta_\text{m} - \delta_\text{r} \right), \\
  \theta'_\text{r} &= \frac{k^2}{4} \delta_\text{r} + \frac{\dd\, \text{ln} \, M_\text{PBH}}{\dd \eta} \frac{\rho_\text{m}}{\rho_\text{r}} \theta_\text{r} ,\\
    \delta'_\text{w} &= -(1+w) \left(\theta_\text{w} + \gamma'/2 \right),\\
     \theta'_\text{w} &= \frac{w}{1+w}k^2 \delta_\text{w} -{\mathcal H }(1-3w)\theta_w.
      \label{pert1}
\end{align}
To solve Eqs.~\eqref{bck1}-\eqref{pert1} simultaneously, we use the following superhorizon initial conditions~\cite{Ma:1995ey},
\begin{align}
\gamma\vert_{\tau=\tau_f}=C(k \tau_f)^2, \hspace{1cm} \delta_w\vert_{\tau=\tau_f}=- \frac{1}{2} (1 + w)C(k \tau_f)^2 \nonumber\\
\theta_w\vert_{\tau=\tau_f}=-\frac{C k^4 \tau_f ^3 w}{2 (4-3 w)}, \hspace{1cm} \epsilon\vert_{\tau=\tau_f}=2C, \nonumber\\
 \delta_r\vert_{\tau=\tau_f}=- \frac{4}{3(1+w)}\delta_w\vert_{\tau=\tau_f}, \hspace{1cm} \delta_m\vert_{\tau=\tau_f}=- \frac{1}{(1+w)}\delta_w\vert_{\tau=\tau_f} , \nonumber\\
\theta_w \vert_{\tau=\tau_f}\approx 0, \hspace{1cm} \theta_r \vert_{\tau=\tau_f}\approx 0, \hspace{1cm} \theta_m=0 \, .
\end{align}
It is important to note that, though the initial values of  $\theta_w $ and $\theta_r$ are non-zero, for the superhorizon perturbations $k\tau \ll 1$ and considering only the linear order of $k\tau$ in the equations, we can take $\theta_w\vert_{\tau=\tau_f}  \approx \theta_r\vert_{\tau=\tau_f} \approx 0$, while we consider  $\theta_m =0$ throughout the evolution. We can connect these synchronous gauge perturbation variables to Bardeen potential of the conformal Newtonian gauge $\Phi$ with,
\begin{align}
\Phi(k,\tau)=\epsilon- \frac{\mathcal{H}\left(  6 \epsilon' + \gamma' \right)}{2k^2}
\end{align}
\begin{figure}[t]
\begin{center}
\includegraphics[height=6.9cm]{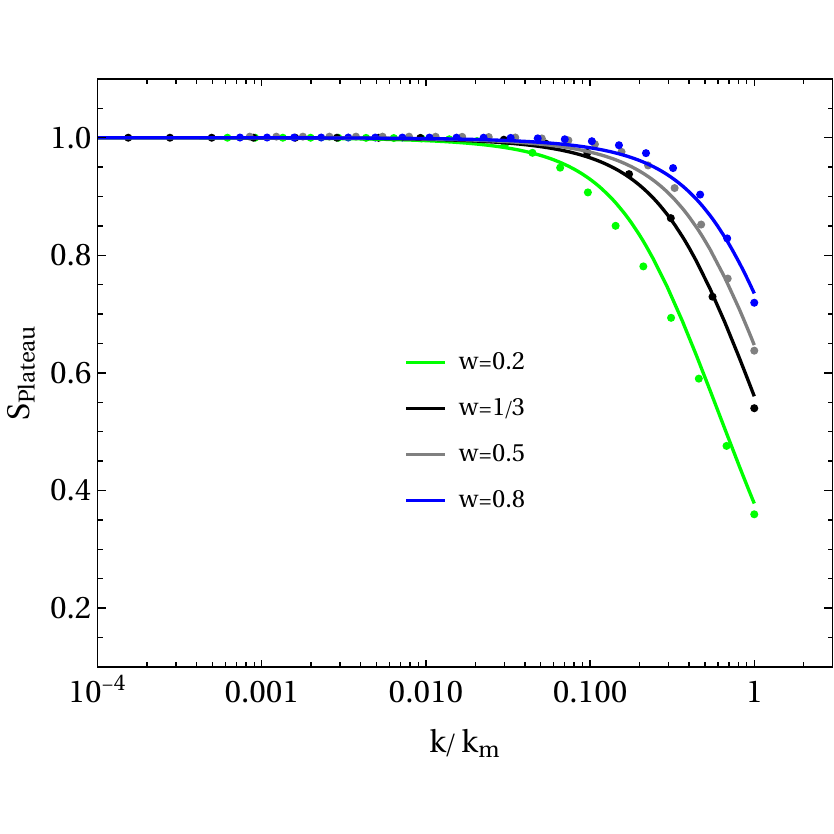}
\caption{The comparison of our adopted interpolation function (Eq.~\eqref{Spl}) and numerical results for $S_{\rm plateau}$ for different values of $w$. }
\label{res-spp}
\end{center}
\end{figure}
At the early stage of PBH domination, after a perturbation mode re-enters the horizon, the first-order scalar perturbation denoted by the gravitational perturbation $\Phi(k)$, suffers a suppression before saturating to a nearly constant plateau value. We denote this suppression as $S_{\rm plateau}$,
\begin{equation}
S_{\rm plateau}(k)=\frac{\Phi(k, \tau_{\rm plateau})}{\Phi(k,\tau_m) } \, .
\end{equation} 
For RD to PBH domination transition, we can solve Eqs.~\eqref{bck1}-\eqref{pert1} to obtain a simple interpolation function for $S_{\rm plateau}$  (BBKS solution)\cite{Bardeen:1985tr}, 
\begin{align}
S_{\rm plateau-RD}(x_{m})=&\;\frac{\log(1 + 0.214 x_{m})}{(0.214 x_{m})}  \times \nonumber \\
&\left(1 + 
    0.355 x_{m} + (1.473 x_{m})^2 + (0.499 x_{m})^3 + (0.612
x_{m})^4 \right)^{-1/4} \, ,
\label{Sp-RD}
\end{align}
where we use the dimentionless variable $x_m \equiv k \tau_m = k/k_m$. We find that it is possible to obtain an approximate generalization to this fitting function by introducing a $w-$dependent shift to $x_{m}$ in the expression of $S_{\rm plateau-RD}$ in Eq.~\eqref{Sp-RD},
 \begin{equation}
 S_{\rm plateau}(w,x_{m})=S_{\rm plateau-RD}(f(w) \times  x_{m}) \, ,
 \label{Spl}
\end{equation}
where 
\begin{equation}
f(w)=-56.4268\, +48.193 w^3+\frac{0.0767357}{w^3}-120.733
   w^2-\frac{1.67478}{w^2}+117.778 w+\frac{14.28}{w}.
\end{equation}
As evident from Fig.~\ref{res-spp}, the numerical results agree reasonably well with this polynomial fitting function. This suppression encoded in $S_{\rm plateau}$ is due to the presence of fluid component with EOS $w$, even after the PBH domination starts. Thus, the suppression is most pronounced for modes that re-enter closest to the PBH domination epoch, and the highest possible suppression is obtained for modes with $k=k_m$ or $x_m=1$. On the other hand, modes that re-enter far away from the PBH domination suffer no suppression, and we can normalize $ S_{\rm plateau} \to 1$ for $x_m \to 0$. It is important to note that in Fig.~\ref{res-spp}, we plot $S_{\rm plateau}$ only for $k \le k_m$. For modes with $k> k_m$ which re-enter the horizon before PBH domination, we assume the relative suppression $\frac{\Phi(k,\tau_{\rm plateau})}{\Phi(k, \tau_m)} \approx \frac{\Phi(k_m,\tau_{\rm plateau})}{\Phi(k_m,\tau_m)}$, in other words for $x_m > 1$ we take $S_{\rm plateau}(w,x_{m}) \approx S_{\rm plateau}(w,1)$.

Moreover, close to the onset of RD, as the PBHs start to lose their mass rapidly, the gravitational potential suffers another suppression which we denote as $S_{\rm ev}$ \cite{Inomata:2019zqy,Inomata:2020lmk}, where,
  \begin{align}
  S_{\rm ev}(k,k_r)\equiv\frac{\Phi(k,\tau_r)}{\Phi(k,\tau_{\rm plateau})}=\left(\frac{\sqrt{6}\,k_r}{k}\right)^{1/3} \, .
  \end{align}
  This second contribution effectively sums up the suppression throughout the PBH domination phase after the ``plateau" value until the perturbation mode decouples from the non-relativistic matter fluid. We can denote the total suppression of the gravitational potential from its initial value to its final value at the onset of RD before the oscillation and decay during RD as $S_{\rm eff}$,
  \begin{align}
S_{\rm eff}(w, k,k_m,k_r)=S_{\rm plateau}(w,x_m\equiv k/k_m) \times S_{\rm ev}(k,k_r)
\end{align}

If we assume the isocurvature-induced adiabatic and inflationary adiabatic components to be uncorrelated, the combined contribution of adiabatic scalar perturbations at the onset of RD can be written as,
\begin{align}
    \mathcal{P}_{\Phi}(k,\tau_r)= \left( \mathcal{P}_{\Phi_{\rm PBH}}(k,\tau_m)+ \mathcal{P}_{\Phi_{\rm infl}}(k,\tau_m) \right) \times \left(S_{\rm eff}(w, k, k_m, k_r)\right)^2
    \, .
\end{align}
\begin{figure}[!t]
\begin{center}
\includegraphics[height=5.5cm]{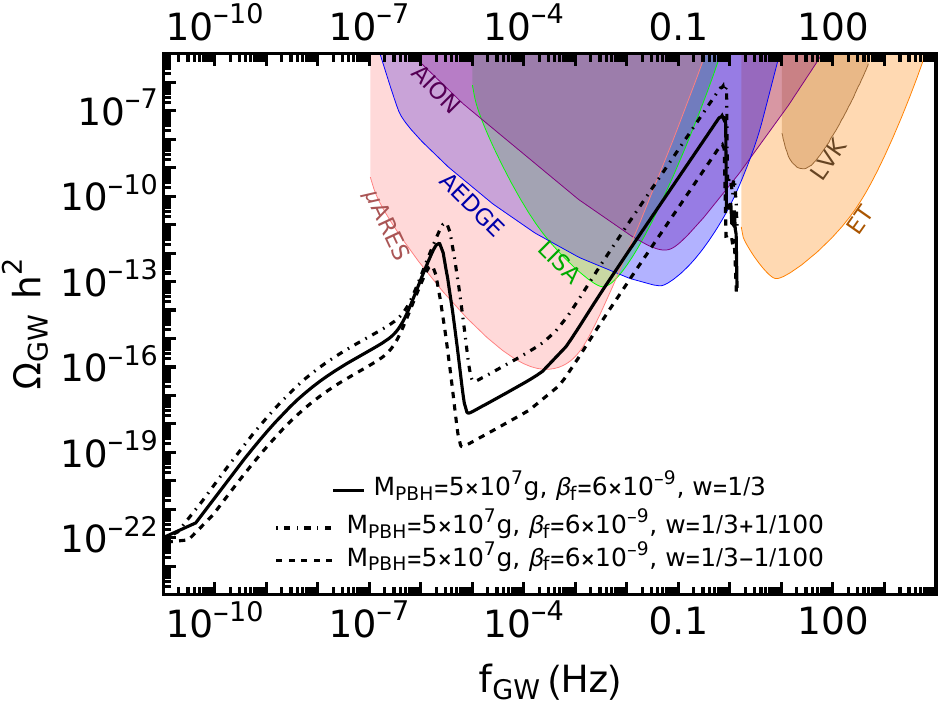}
\hspace{0.1cm}
\includegraphics[height=5.5cm]{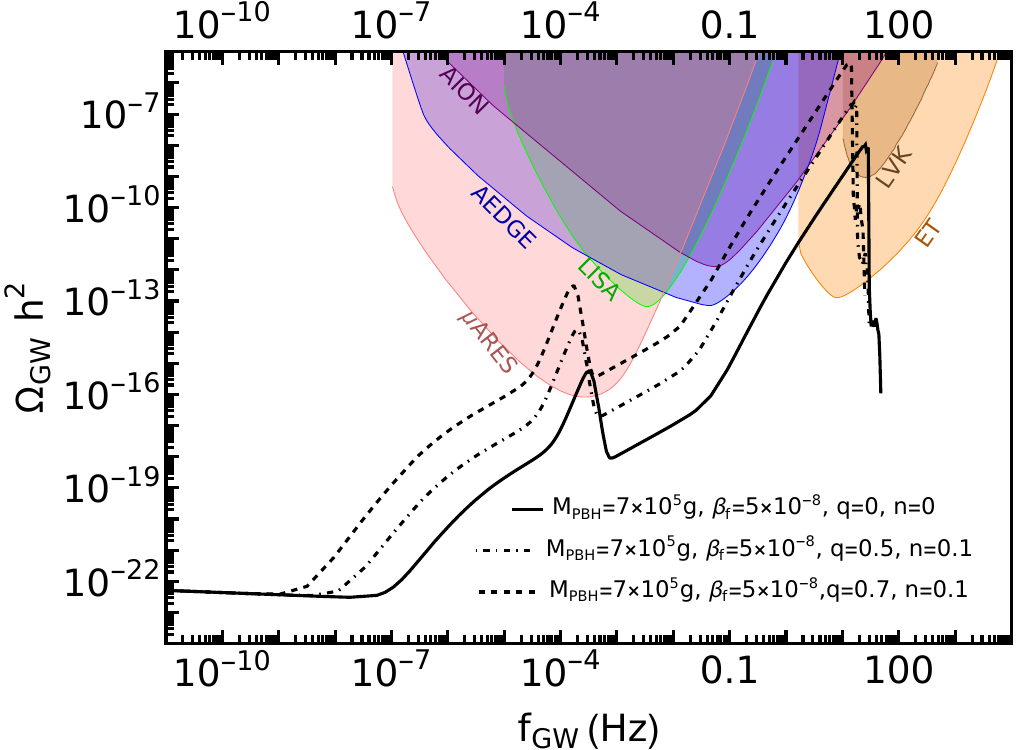}
\vskip 6pt
\caption{\textbf{Left panel:} The dimensionless spectral energy density of the induced stochastic GW background (ISGWB) is plotted as a function of frequency for the initial PBH mass $\MPBH =5 \times 10^7 {\rm g}$ and $\beta_f= 6 \times 10^{-9}$ with different reheating histories. Note that we use a very tiny deviation in the EOS of the universe before PBH domination, which leads to significant changes in the ISGWB spectra. \textbf{Right panel:} We have shown the effect of memory burden on the ISGWB spectrum for PBH parameters $\MPBH =7 \times 10^5 {\rm g}$ and $\beta_f=5 \times 10^{-8}$ and compared it with the standard case. To do that, we assume standard RD before PBH domination.}
\label{res-beta1}
\end{center}
\end{figure}

Now we can use this $ \mathcal{P}_{\Phi}(k,\tau_r)$, to get the power spectra of  comoving curvature perturbation $ \mathcal{P}_{\mathcal R}(k)$ at the very start of RD and estimate energy density per logarithmic $k$ interval $\Omega_{\rm GW}(\tau,k)$ for the second-order SGWB sourced by the first-order scalar perturbations, 
\begin{align}
\label{omega_uv}
\!\!
\Omega_{\rm GW}(\tau,k) =\frac{1}{6}  \int_{0}^{\infty}dv \int_{|1-v|}^{1+v} du
\left(\frac{4 v^2-(1+v^2-u^2)^2}{4 u v}\right)^2
\times \overline {\cal I}^2(v,u,x) {\cal P}_{\cal R}(kv){\mathcal P}_{\mathcal R}(ku) 
\end{align}
For the resonant SGWB generated during RD,  $\overline{\cal I}^2 \equiv \overline{\cal I}_{\rm RD}^2 $ comes from the evolution of first-order scalar perturbation modes during RD. Due to early matter domination (eMD) before the late RD phase, the expression of $\overline{\cal I}_{\rm RD}^2 $ differs from what one would expect in the case of standard RD starting just after inflation. Considering a finite duration of PBH-dominated era ($\tau_{m} \ll \tau_{r}$), we get the scale factor during RD $a_{\rm RD} \propto (\tau - \tau_r/2)$~\cite{Bhaumik:2022pil}. This allows us to use the expression for  $\overline {\cal I}_{\rm RD}^2$  derived in Appendix A of our earlier work~\cite{Bhaumik:2020dor}. Here, we also assume the PBH peculiar velocities have negligible contributions. We can then express the present day ($\tau=\tau_0$) value of $\Omega_{\rm GW}$,
\begin{equation}
\Omega_{\rm GW}^{\rm res}(\tau_0,k) =  c_g ~\Omega_{r,0} ~\Omega_{\rm GW}^{\rm res}(\tau_l,k)\, ,
\end{equation}
where the present radiation energy density is $\Omega_{r,0}$ and $c_g \approx 0.4$ when relativistic degrees of freedom are taken $\sim 106.7$, and $\tau_l$ corresponds to the conformal time during RD when the source term stops to contribute to the kernel and the kernel term saturates~\cite{Espinosa:2018eve}.

We plot the resultant SGWB for different reheating histories before PBH domination in the left panel of Fig.~\ref{res-beta1}
along with the projected sensitivities of
LIGO/Virgo/KAGRA (LVK)~\cite{LIGOScientific:2014pky, LIGOScientific:2016fpe, LIGOScientific:2019vic} at the end of its operation, the Einstein Telescope (ET)~\cite{Punturo:2010zz, Hild:2010id}, LISA~\cite{Bartolo:2016ami, Caprini:2019pxz, LISACosmologyWorkingGroup:2022jok}, and atom interferometry experiments AION~\cite{Badurina:2019hst} and AEDGE~\cite{AEDGE:2019nxb} and
 $\mu$ARES\cite{Sesana:2019vho}.
The low-frequency peaks are associated with the first-order inflationary scalar perturbations, which source the tensor perturbations at the second-order. We take the cutoff of inflationary adiabatic perturbation at $k=k_m$ because modes with $k>k_m$ re-enter during the reheating phase and get significantly suppressed. In the poltergeist mechanism, the cutoff scales correspond to the peak frequencies.  This transition point $k_m$ depends on the background EOS $w$ during which PBHs form. Increasing $w$ indicates a faster redshift of the inflaton field, and consequently, the transition point shifts closer to the end of inflation, which is also visible in Fig.~\ref{res-beta1} that increasing $w$ shifted the peak frequency towards a higher value closer to inflation end. On the other hand, the high-frequency peaks in Fig.~\ref{res-beta1} are from the second-order tensor perturbations sourced by PBH density fluctuations induced adiabatic perturbations.  The peak value of the second peak is associated with the cutoff scale $k_{\rm UV}$ of PBH density perturbations. Thus, as shown in Eq.~\eqref{Eq:Kuv}, $k_{\rm UV}$ and the position of the second peak bears a weak dependence on $w$ and are nearly identical.

To summarize, the SGWB spectrum depends on the background EOS$w$ wherein PBHs form. The lifetime of PBHs does not depend on the value of $w$. However, changing $w$
 determines how quickly the universe reaches PBH domination. This changes the duration of PBH domination and, as a consequence, has an impact on the frequency and amplitude of the SGWB peaks.
 The value of $w$ determines how PBH density fluctuation $\delta_m$ or isocurvature perturbations evolve during $w$-domination, as evident from Mészáros equation \eqref{isoeq}. At the same time, the changes in $w$ modifies the amount of suppression the combined adiabatic perturbation $\Phi(k,\tau)$ suffers during PBH domination, which we compute through the $w$ dependence of $S_{\rm plateau}(w,x_m)$. These manyfold aspects contribute to a significant modification in the SGWB even for a very small change in the EOS $w$ before PBH domination, as evident from the panel of Fig.~\ref{res-beta1}.
 
\section{The memory burden effect}
\label{sec-mbe}

In this section, we shall first briefly review the results derived from the semiclassical formulation of Hawking evaporation and then discuss the modifications in the mass dissipation rate and lifetime of the PBHs if one takes into account the quantum corrections in Hawking evaporation, such as the memory burden effect. We consider a chargeless non-rotating PBH poulation with a monochromatic mass range. For initial mass $M_{\rm PBH} $, the horizon temperature of these PBHs can be written as~\cite{Hawking:1975vcx,Page:1976df},
\begin{equation}
T_{\rm BH} = \frac{\Mpl^2}{%8 \pi
M_{\rm PBH}} \, ,
\end{equation}
and the rate of mass loss  per unit time as,
\begin{equation}
\frac{d M}{d t} = -\frac{ \pi~ \mathcal{G}~ g_{*, H} \Mpl^4}{480  M_{\rm PBH}^2} \, .
\end{equation}
The semiclassical estimation of the PBH lifetime in terms of physical time, $\Delta t_{\rm 0}$ can be written as,
\begin{equation}
\label{tevp}
    \Delta t_{\rm 0}=\frac{160\,  M_{\rm PBH}^3}{\pi~ \mathcal{G}~ \overline{g_{*,H}} ~\Mpl^4 } \, ,
\end{equation}
where $\mathcal{G} \approx 3.8$ is the effective graybody factor,  $g_{\star, H}$  is the degrees-of-freedom at temperature $T_{\rm BH}$ and $\overline{g_{*,H}}$ denotes the average over PBH lifetime \cite{Hooper:2020otu,MacGibbon:1991tj}. Considering only the standard model particles and limiting ourselves to $M_{\rm PBH} \le 10^9 {\rm g}$, we can take $\overline{g_{*,H}} \approx 108$.

Now, let's explore how the memory burden effect on PBH evaporation modifies the mass dissipation rate. Before the memory burden effect starts to accumulate, we can always assume the very initial stage of Hawking evaporation to be accurately described by purely semiclassical methods. Let us assume that this initial phase lasts up to the time when PBH mass decays to $q  M_{\rm PBH}$ with initial PBH mass being $ M_{\rm PBH}$. Thus, the time elapsed during this initial phase can be written as $t_{\rm q}=(1-q^3)\Delta t_{0}$. Once the memory burden effect accumulates, the PBH evaporation process must deviate from the self-similar semi-classical decay, and after $t=t_{\rm q}$ there are two fates of the PBH, one can infer: (i) PBH can disintegrate very quickly due to classical instabilities, or (ii) The memory burden effects can further stabilize the black hole against Hawking evaporation, thus prolonging its lifetime \cite{Dvali:2020wft}. We only pursue the second possibility in this work, and following the formalism of \cite{Dvali:2020wft, Alexandre:2024nuo, Thoss:2024hsr}, we can express the diminished mass loss rate of PBH as
\bea 
\f{\d M}{\d t}=-\frac{ \pi~ \mathcal{G}~ g_{*, H}\Mpl^4}{480M^2}  
\begin{cases}
1& \text{for}~M>q\Min\\
&\\
S^{-n}(M)& \text{for}~M<q\Min
\end{cases}\, .
\label{eq:ti}
\eea  
 Here $S(M) = M^2/(2\Mpl^2)$ denotes the mass-dependent entropy of the PBH, and the parameter $n$ signifies the strength of the memory burden effects. Higher the value of $n$, we expect the memory burden stabilization effects to be stronger, and we recover the semiclassical results in the $n \to 0$ limit. Since we don't know from which point the memory burden effect starts or the exact strength of memory burden effects, we assume both $q$ and $n$ as free parameters. Smaller values of $q$ imply the memory burden effects to become significant later, and in the $q\to 0$ limit, we assume the whole evaporation history of PBH to be described by semiclassical evolution.

The memory burden effect alters the PBH lifetime and thereby changes the cosmic history in the case of an ultra-low mass PBH domination scenario. If we assume the PBH lifetime without memory burden effect for Schwarzschild black holes as $\Delta t_{0}$, it is possible to write the modified lifetime $\Delta t$ due to the memory burden effect as,
\bea 
\Delta t = \Delta t_{0} \mathcal{F}_{\rm MBE}(q,n)
\label{lift}
\eea  
where 
\bea 
\mathcal{F}_{\rm MBE}(q,n)= (1 - q^3)+ \frac{3q^{3+2n}S^n(\Min)}{3+2n}
\label{mylife}
\eea
This change in the PBH lifetime\footnote{Note that, our expression differs from \cite{Balaji:2024hpu} by the initial factor $(1 - q^3)$ which refers to the time elapsed while PBH evaporation still follows semiclassical dynamics. It is evident from Eq.~\eqref{mylife} that we recover the semiclassical value $\mathcal{F}_{\rm MBE}(q,n)=1$ if we take either $q\to 0$ or $n\to 0$ limit. In the limit, $n \gg 0$ and $M_{\rm PBH} \gg M_{pl}$ used in \cite{Balaji:2024hpu}, we can neglect the initial duration before the memory burden effect in comparison with the total PBH lifetime, and our results concur.} also impacts upon the wavenumbers associated with the SGWB generation. We can express the effect in terms of $\mathcal{F}_{\rm MBE}$,
\begin{eqnarray}
k_{r} &=& k_{r0}  [\mathcal{F}_{\rm MBE}(q,n)]^{-1/2} , \\
k_{m} &=&k_{m0} C^{2/3} [\mathcal{F}_{\rm MBE}(q,n)]^{-1/6}\, ,  \\
k_{f}&=&k_{f0} C^{-1/3} [\mathcal{F}_{\rm MBE}(q,n)]^{-1/6}\, , \\
k_{\rm UV}&=&k_{\rm UV0} C^{-1/3} [\mathcal{F}_{\rm MBE}(q,n)]^{-1/6} \, ,
 \label{wavesmbR}
\end{eqnarray}
where the "0" in subscript refers to the wavenumber values obtained in semiclassical approximation without taking into account the memory burden effect. Depending on the initial abundance of the PBHs $\beta_f$ and the value of $q$, we find the correction factor $C$ for four distinct possibilities (cases),
\begin{enumerate}
\item The memory burden effect starts far after PBH domination, and at the time of PBH domination, the change in the mass of the PBH from its initial value is negligible. \label{case1}
\item The memory burden effect starts soon after PBH domination and at the time of PBH domination the change in the mass of the PBH from its initial value is not negligible.\label{case2}
\item The memory burden effect starts before PBH domination, and at the time of PBH domination, the mass of the PBH is $q$ times its initial value. \label{case3}
\item  The memory burden effect starts after PBH domination, and for $q < 0.5$, there is a possibility of intermediate RD between two PBH-dominated phases. \label{case4}
\end{enumerate}

For simplicity, we limit ourselves to the first three cases and work with appropriate values of $q$ to avoid the intermediate RD phase. 

If we assume that the PBH mass does not change significantly before the PBH domination, we can take $C=1$ (case \ref{case1}). The assumption that PBHs do not lose a significant fraction of their mass until their domination always holds in the case of semiclassical PBH evaporation because after PBH loses a non-negligible fraction of mass, PBH evaporation rate is so high that it rules out the possibility of PBH domination. However, when we consider the memory burden effect, this assumption may break down for certain values of $\beta_f$, typically if $\beta_f$ is such that PBH dominates very close to its semiclassical lifetime or afterwards. In such a scenario, it becomes necessary to take the correction factor $C \neq 1$.

If we assume that PBH of initial mass $M_{\rm PBH}$ loses its mass significantly at scale factor $a=a_y$
and the mass of the PBH at the time of PBH domination is $y M_{\rm PBH}$, equating the PBH and radiation energy density at eRD to MD transition we get,
\bea
y \beta_f = \left(\frac{a_f}{a_m} \right)+(1-y)\beta_f \left(\frac{a_y}{a_m} \right) \, ,
\eea
which leads to 
\bea
\left(\frac{a_f}{a_m} \right) = y \beta_f-(1-y)\beta_f \left(\frac{a_y}{a_m} \right)  \,.
\label{mark1}
\eea
In the scenario when for $q<y<1$, and PBH dominates before memory burden effect starts (case  \ref{case2}), $a_y \approx a_m$ and Eq.~\eqref{mark1} simplifies to
\bea
\left(\frac{a_f}{a_m} \right) = (2y-1) \beta_f \, .
\label{mark2}
\eea
For $q<y<1$, the fraction of mass  at the time of PBH domination is $y$, which can be derived by solving,
\bea
1 - 4 y + 4 y^2 - y^3 + 4 y^4 - 4 y^5=\left(\frac{\beta_c}{\beta_f}\right)^2 \, ,
\eea
where $\beta_c =\sqrt{\frac{\mathcal{G}\,g_{*, H}\Mpl^2}{1280\gamma}}\frac{\Mpl}{M_{\rm PBH}}$. This leads to the correction factor 
\bea C=(2y-1) \, .
\label{mark3}
\eea
It is interesting to note that we recover the no mass loss limit of Eq.~\eqref{mark3} and \eqref{mark2} for $y\to 1$.

On the other hand, if PBH domination happens after the memory burden effects start to kick in (case  \ref{case3}), we can take $y\to q$, $a_y\to a_q$ and Eq.~\eqref{mark1} can be written as
\bea
\left(\frac{a_f}{a_m} \right) = \frac{q \beta_f}{1+(1-q)\beta_f \left(\frac{a_q}{a_f}\right)}\, ,
\eea
where 
\bea
\left(\frac{a_q}{a_f}\right)=\frac{ \sqrt{1-q^3}}{\beta_c} \,.
\eea
This leads to a correction factor,
\bea
C= \frac{q}{1+(1-q)\sqrt{1-q^3}\left(\frac{ \beta_f}{\beta_c}\right)}\, .
\eea
Thus, the correction factor $C$ can be summarized as\footnote{Our formulas for the correction factor differ from ref. \cite{Balaji:2024hpu}, as we consider broader possibilities, while the correction factor used in \cite{Balaji:2024hpu} can be obtained in a particular  limit ($a_m \gg a_q$) of our case  \ref{case3} in Eq.~\eqref{sumup}.},
\begin{equation}
C=
\begin{cases}
1 \quad\quad\quad\quad\quad\quad\quad\quad\quad\,\,\text{for case  \ref{case1}},\\
&\\
2y-1\quad\quad\quad\quad\quad\quad\quad\text{for case  \ref{case2}},\\
&\\
\frac{q}{1+(1-q)\sqrt{1-q^3}\left(\frac{ \beta_f}{\beta_c}\right)}\quad\quad\text{for case  \ref{case3}},
\end{cases}
\label{sumup}
\end{equation}
We can show that when PBH domination starts before the memory burden effect kicks, in we have $q\leq0.5+(a_m/a_q)\sim 0.5$ for $a_m \ll a_q$, there will be another intermediate RD phase arising during PBH domination (case  \ref{case4}). In this case, our analysis is not applicable, and we do not consider this part of the parameter space.

On the other hand, for case~\ref{case2}, we always expect $y>0.5$ because $y=0.5$ refers to the situation when the radiation component coming from PBH evaporation equals the PBH component, and the PBH component would certainly dominate far before that happens, as evident from Eq.~(\ref{mark2}). However, for case \ref{case3}, even with $q\leq0.5$, we can expect PBHs to dominate later on. Note that we assume instantaneous production of PBH decay products. However rapid the pace is, in a more rigorous treatment, the accumulation of PBH decay products takes a finite amount of time. The correction due to this particular factor leads to $q < 0.5 $ as the limiting value for the occurrence of case \ref{case4}, which can be obtained numerically, and would depend on various factors like the proximity of PBH domination, initial PBH spin and etc.

Once the dependence of these wavenumbers on the memory burden effect is understood, it is straightforward to follow the same formalism described in the previous section to estimate the initial suppression $S_{\rm plateau}$ with $w=1/3$, but in the case of memory burden effect we need to modify the suppression of $\Phi(k, \tau)$ from plateau value to the value at the start of RD. Depending on the situation we are considering, there can be two distinguishable suppression effects, one during the start of memory burden effect at $M=q M_{\rm PBH}$ denoted by $S_q \equiv \Phi(k, \tau_{q})/\Phi(k, \tau_{\rm plateau})$, where $\tau_q$ is the conformal time when PBH mass reaches $M=q M_{\rm PBH}$. The second suppression occurs during the final evaporation of PBHs, which we denote by $S_{\rm ev}$. The suppression of $\Phi(k, \tau)$ during the start of the memory burden effect is simply proportional to the change in the PBH mass and thus for different cases, we can express $S_q$ as,

\begin{equation}
S_q=
\begin{cases}
q\quad\quad\quad\quad\quad\text{for case \ref{case1}},\\
&\\
\frac{q}{y}\!\quad\quad\quad\quad\quad\text{for case \ref{case2}},\\
&\\
1\quad\quad\quad\quad\quad\text{for case \ref{case3}} \, .
\end{cases}
\label{Sq}
\end{equation}
It is evident that as in the case \ref{case3}, we consider PBH domination to start after the memory burden effect, we expect no suppression factor during the start of memory burden effect, and $S_q$ must be unity. On the other hand,  the suppression factor $S_{\rm ev}$ before $\Phi(k, \tau)$ decouples from the matter fluids can be written as,
\begin{equation}
S_{\rm ev}=\left( \frac{\sqrt{3}\sqrt{3\alpha-1}}{\mu\alpha} \frac{k_r}{k} \right)^{\frac{1}{3\alpha}} \, ,
\end{equation}
where,
\begin{equation}
\alpha = 1+ \frac{2n}{3} \, ,\nonumber \quad
\mu = \frac{\lambda}{(1-q^3)(2n+3)+\lambda} , \nonumber \quad
\lambda =3q^{2n+3}{{\cal S}_f}^n \, ,
 \end{equation}
where ${\cal S}_f$ denotes the entropy of the PBHs at their formation. Evidently, in this case, the total suppression $S_{\rm eff}$ can be expressed as,
  \begin{align}
S_{\rm eff}(k,k_m,k_r,q,n, M_{\rm PBH})=S_{\rm plateau}\times S_q \times S_{\rm ev}
\end{align}
Once the wavenumber dependencies are resolved and the amount of suppression of adiabatic perturbation during PBH domination is understood, we can estimate the
scalar perturbation profile at the end of PBH domination and, thus, the resonant SGWB spectra generated at the onset of RD.

As we already mentioned, the memory burden effect delays the PBH evaporation process. Thus, for PBH with a specific mass and abundance, the memory burden effect elongates the duration of the PBH domination compared to the standard Hawking evaporation case. Since the result we described for SGWB strictly depends on the duration of the PBH domination, it has been imprinted on the spectrum of SGWB through two memory burden parameters ($n$, $q$). As we can see in the right panel of Fig.~\ref{res-beta1}, the memory burden effect impacts both the peak frequency and the amplitude of the SGWB.

\section{Degeneracy of the memory burden effect and non-standard reheating}
\label{sec-deg}

As we focus on probing the ultra-low mass PBH domination scenarios by detecting doubly peaked GW spectrum, understanding the possible degeneracies becomes increasingly important. Interestingly, if we consider just the variation of reheating history isolated from the memory burden effect, there is no degeneracy in the SGWB spectra. In other words, two different combinations of parameters [$w$, $\beta_f$, $M_{\rm PBH}$] can not reproduce the same SGWB spectral shape for any of the two peaks we consider. Similarly, though memory burden parameters $q$ and $n$ have internal degeneracy with each other, the overall memory burden effect characterized by any two values of $n$ and $q$ does not lead to any degeneracy with the variation of PBH parameters $\beta_f$ and  $M_{\rm PBH}$.
Therefore, when memory burden effects and variation of reheating histories are considered separately, the detection of any of the resonant SGWB peaks could be used to determine the parameters uniquely. However, there can be a degeneracy when these two effects are considered together. 
\begin{figure}[t!]
\begin{center}
\includegraphics[height=8cm]{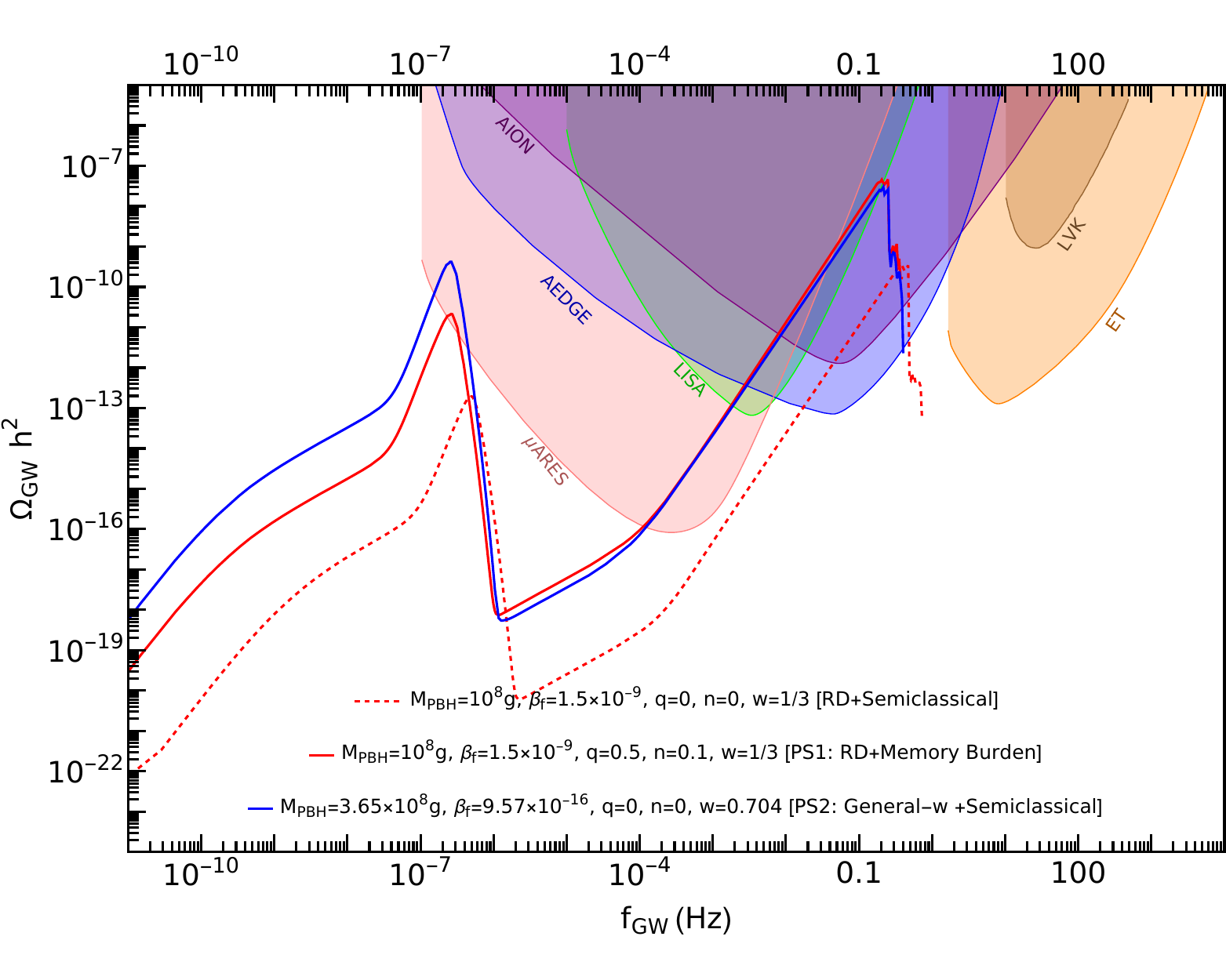}
\vskip 2pt
\caption{We plot the ISGWB spectral energy density with (solid red: $PS1$) and without (dashed red) memory burden effect and also plot SGWB for the degenerate set of parameters (solid blue: $PS2$) with different background equations of state before PBH domination. Evidently, in these cases of the high-frequency peak or the PBH density fluctuation peak, PBH formation during reheating and evaporation due to standard Hawking evaporation can mimic the effect of memory burden in the case of PBH formation during RD. However, this degeneracy is broken if we can simultaneously also detect the amplitude of the first peak or the inflationary adiabatic SGWB peak.}
\label{res-beta2}
\end{center}
\end{figure}

We adopt a rather straightforward approach to check the existence of any degeneracy. We consider only three quantities: the conformal time when PBHs evaporate and resonant SGWB is generated ($\equiv 1/k_r$), the mean comoving distance between two PBHs ($\equiv 1/k_{\rm UV}$) and the duration of PBH domination ($\tau_{\rm rat}$) and check if it is possible to get unique values of these three quantities for any two sets of parameters;
\begin{align}
&\text{Parameter set 1 }: PS1 \equiv \{ M_{\rm PBH1}, \beta_{f1}, w_1=1/3, q_1, n_1 \} \nonumber \\
&\text{Parameter set 2 }: PS2 \equiv \{ M_{\rm PBH2}, \beta_{f2}, w_2, q_2=0, n_2=0 \} \nonumber 
\end{align}
We consider case \ref{case1} or $C=1$ to find the degenerate parameter sets in a memory burden to mimic general $w$ results and use three conditions,

\begin{enumerate}
\item We shift the mass of the PBH to cancel out the MBE in the expression of $k_r$ by solving
\bea
k_r \vert_{PS1}=k_r\vert_{PS2} \, .
\label{con1}
\eea
Since $k_r$ has no dependence on reheating history EOS $w$ or PBH abundance $\beta_f$, the redefined mass $M_{\rm PBH2}$ is only a function of the memory burden parameters. The lifetime of the PBHs sets the comoving time of PBH evaporation and also fixes the physical time of PBH evaporation.

\item We use the redefined mass to get the optimal equation of state for the reheating $w_2$ that gives $k_{\rm UV}$ identical with MBE,
\bea
k_{\rm UV} \vert_{PS1}=k_{UV} \vert_{PS2} \,
\label{con2}
\eea
Here, we use the fact that $k_{\rm UV}$ is completely independent of PBH abundance $\beta_f$. This property also ensures that the only memory burden effect or effects of the variation of reheating history is not degenerate with PBH parameters, $\beta_f$ and $M_{\rm PBH}$.

\item Finally, we shift the PBH abundance $\beta_f$ to obtain the same duration of PBH domination for both with and without memory burden effect,
\bea
\tau_{\rm rat} \vert_{PS1} =\tau_{\rm rat} \vert_{PS2} \, .
\label{con3}
\eea
\end{enumerate}
Thus, it is evident that we can always obtain the parameter set [$w_2$, $M_{\rm PBH2}$, $\beta_{f2}$] of non-standard reheating history without considering the memory burden effect ($n \to 0$, $q \to 0$) which can reproduce the memory burden effect ($q\neq 0$, $n\neq0$)  modifications to the useful comoving wavenumber sets for standard reheating history [$w=1/3$, $M_{\rm PBH1}$, $\beta_{f1}$], and vice versa. 
\begin{figure*}[t!]
\begin{center}
\includegraphics[width=6.5cm, height=6.7cm]{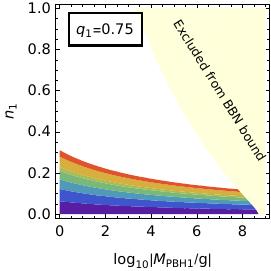}
\includegraphics[width=7.9cm, height=6.7cm]{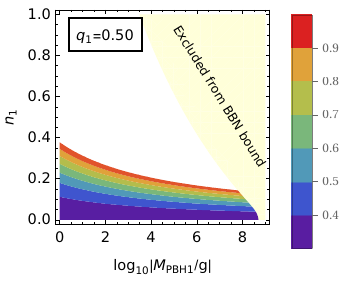}
\vskip 6pt
\caption{
The rainbow-coloured contours show the values of the EOS $w_2$ of $PS2$, required to mimic the memory burden effect as a function of memory burden parameter $n_1$ and PBH parameter $M_{\rm PBH1}$  for a few fixed values of the other memory burden parameter $q_1$ of $PS1$, which directly follows from Eq.~\eqref{con1} and \eqref{con2}. Colours in the contours refer to different values of $w_2$ as listed in the colour bar on the right. As we set the range of this plot $0<w_2<1$, we can see that for a fixed value of $q_1$, it is not possible to cover the full range of $n_1$ with the variation of reheating history parameter $w_2$ and PBH mass $M_{\rm PBH1}$. It is clear, however, that variations of $n_1$, $q_1$, and $M_{\rm PBH1}$ of $PS1$ can mimic the effects of any values of $w_2$ of $PS2$. We also plot the light yellow shaded region where the required PBH mass of $PS2$, $M_{\rm PBH2} > 5\times 10^8 {\rm g}$, and thus this part of the parameter space is excluded from BBN bound.
}
\label{3mim}
\end{center}
\end{figure*}
We illustrate in Fig.~\ref{res-beta2} what kind of SGWB spectra this degenerate set of parameters leads to.  The solid red and dashed red lines show the spectra associated with a given population of PBHs with and without including the impact of the memory burden effect. The solid blue line shows the spectrum associated with a different PBH population in a non-standard reheating scenario. This shows that with $PS1$ (solid red) and $PS2$ (solid blue), we get an identical second peak of PBH density fluctuation peak, while the first peak or inflationary adiabatic peak is slightly different.  To mimic the effect of $w$ through the memory burden effect for the second peak, we need to change both PBH parameters $\Min$ and $\beta_{\rm f}$. Fig.~\ref{3mim} shows that the degeneracy is not complete because the equation of state $w_2$ during reheating can only take values $0<w_2<1$~\cite{Allahverdi:2020bys}. While choosing different values of $q$ can effectively allow one to mimic the whole range of reheating possibilities, nonstandard reheating can only mimic a limited part of the memory burden effect parameter space.

It is evident that the first and second equality conditions Eq.~\eqref{con1} and \eqref{con2} are enough to check the existence of degeneracy; the third condition Eq.~\eqref{con3} only indicates that for a degenerate set of parameters in non-standard reheating history without memory burden effect that mimics another set of parameters with memory burden effect, the PBH abundances in these two setups must be different to ensure that they will have identical duration of PBH domination. This guarantees that in these degenerate setups, we get PBH evaporation at the same time (from Eq.~\eqref{con1} ), they also dominate for the same time (from Eq.~\eqref{con3}), and they have the same mean separation (from \eqref{con2}), but they do not form at the same time ($k_{f1} \neq k_{f2}$). Even with these three interesting degenerate values, the first peak or inflationary adiabatic peak differs from each other as the suppression factor $S_{plateau}$ has a formidable dependence on $w$. On the other hand, in the case of the second peak, the combined effects lead to nearly identical peaks. The first peak or inflationary adiabatic peak occurs at the same frequency associated with identical $k_m$ values, the amplitude is quite different for $PS1$ and $PS2$, and thus can be used to break the degeneracy.

On the other hand, we can also consider some particular formation mechanisms for these ultra-low mass PBHs. Amplified inflationary scalar power spectra can lead to PBH formation (e.g. see \cite{Ezquiaga:2017fvi, Garcia-Bellido:2017mdw, Kannike:2017bxn, Hertzberg:2017dkh, 
Cicoli:2018asa, 
%Kamenshchik:2018sig, 
%Dimopoulos:2019wew,
Bhaumik:2019tvl,Ballesteros:2020qam, Ragavendra:2020sop,Ragavendra:2023ret} for formation during RD, and \cite{Bhattacharya:2019bvk, 
Harigaya:2023pmw,Liu:2023pau,Domenech:2024rks,Maity:2024odg} for formation during $w$-domination), and the detection of second-order SGWB sourced by these amplified first-order scalar spectra can be used to break the degeneracy. Irrespective of inflationary models or the resultant shape of the power spectra, these two scenarios will produce PBH formation peaks in SGWB spectra at two very different frequencies. Unfortunately, the amplification in SGWB  spectra from the formation of these ultra-low mass PBH will peak at very high frequencies ($\ge 10^{3}\, {\rm Hz}$), which are beyond the reach of present or near future GW probes. A similar difficulty arises if we propose to break the degeneracy using the detection of an SGWB generated from the emitted gravitons during Hawking evaporation from these two cases. We can expect the Hawking evaporation spectra for spinning particles to be quite different between the setups; however, these SGWB spectra will also fall into very high frequencies, detectable in the future~\cite{Aggarwal:2020olq,Goryachev:2021zzn, Domcke:2022rgu,Bringmann:2023gba, Kondo:2024fvd, Valero:2024ncz,Carney:2024zzk,Domcke:2024mfu}.

Another promising path to break this degeneracy can be through indirect probes such as the detection of Baryogenesis \cite{Hooper:2019gtx,Hook:2014mla,Fujita:2014hha,Hamada:2016jnq, Morrison:2018xla,Hooper:2020otu, Perez-Gonzalez:2020vnz,Smyth:2021lkn, Ambrosone:2021lsx,Calabrese:2023key,Calabrese:2023bxz,Gehrman:2022imk} or dark matter generated from ultra-low mass PBH evaporation~\cite{Morrison:2018xla, Gondolo:2020uqv, Bernal:2020bjf, Green:1999yh, Khlopov:2004tn, Dai:2009hx,Allahverdi:2017sks, Lennon:2017tqq, Masina:2020xhk, Baldes:2020nuv,Hooper:2019gtx,Bhaumik:2022zdd,Cheek:2021odj,Cheek:2021cfe,Bernal:2022oha,Borah:2022iym, Haque:2024eyh,Haque:2024cdh} or the detection of extra relativistic degrees of freedom from dark radiation due to Hawking evaporation of these PBHs~\cite{Hooper:2019gtx,Bhaumik:2022zdd,Barman:2024iht,Lunardini:2019zob,Keith:2020jww,Cai:2020kfq,Masina:2021zpu,Arbey:2021ysg,Baker:2021btk,Calza:2021czr,Auffinger:2022khh,Thoss:2024hsr} in future CMB probes like CMB-HD~\cite{CMB-HD:2022bsz} and CMB-Bharat~\cite{CMBbharat:01}.

\section{Conclusion and discussion}
\label{sec-cd}
In this work, we have considered SGWB generation from a population of ultra-low mass non-spinning PBHs with a monochromatic mass range, which dominates the universe for a short period before BBN, for a non-standard reheating history before PBH domination and for memory burden effect modifications to PBH evaporation. We summarise below the various interesting aspects of our findings:
\begin{itemize}
\item
A non-standard reheating history before PBH domination can significantly alter the expansion of the universe between the time of the formation of PBHs and the moment they dominate the universe. Change in $w$ not only determines the duration of evolution for isocurvature perturbation during $w$-domination but also how it evolves. The suppression effect ($S_{\rm plateau}$) in adiabatic scalar perturbation during PBH domination also depends on $w$ as this effect is primarily due to the presence of residual $w$-fluid after the  PBH domination. These combined effects significantly alter the resonant SGWB that we consider in the left panel of Fig.~\ref{res-beta1}.

\item The memory burden effect modifies the PBH lifetime and leads to a modification in the suppression of adiabatic scalar perturbation ($S_{\rm ev}$) around the time of PBH evaporation, and thus leads to very different SGWB signatures compared to the semiclassical predictions, as shown in the right panel of Fig.~\ref{res-beta1}.

\item Both these effects lead to significant changes in the corresponding SGWB spectra. Here, we find an interesting degeneracy between these two effects for the PBH density fluctuation-induced SGWB peak. From Fig.~\ref{res-beta2}, we can see the two different sets of parameters, one with memory burden effect and another with general-$w$ domination before PBH domination, lead to nearly identical SGWB peak for the PBH density fluctuation induced high-frequency peak.

\item Due to the constraints on the value of the EOS parameter $0<w<1$, the reheating history can mimic only a part of the memory burden parameter space; however, the memory burden effect can mimic any variation of the standard reheating history, as evident from Fig.~\ref{3mim}.

\item We find no degeneracy in either non-standard reheating history or memory burden effect if we consider any of them individually. However, when including both these possibilities, this degeneracy would prevent the unique identification of underlying parameters from future detection of SGWB peak from the PBH density fluctuations. We conclude that any indications of deviations from standard reheating history in the PBH density fluctuation peak or second peak in the SGWB can also be explained by memory burden effects, but if we can also simultaneously detect the inflationary adiabatic peak, this degeneracy can be broken, and the memory burden parameters or the general$-w$ domination parameters can be identified uniquely.

\item While our work focuses on this degeneracy with a particularly simple approach in mind, it is also possible to find other kinds of degeneracies in the SGWB spectra (e.g. the degeneracy in the first peak rather than in the second peak) if one approaches the question of degeneracy with a thorough parameter space scan. Our approach is motivated to emphasize the possible existence of this degeneracy, not so much to state that this is the only possible degeneracy.

\end{itemize}
Future studies of the memory burden effect can shed more light on the preferred values of $q$ and $n$ and thus can help to identify the reheating history between the end of inflation and PBH domination through doubly peaked SGWB detection. Other ways of breaking this degeneracy with SGWB involve the SGWB spectra from PBH formation or graviton emission from Hawking evaporation, which would involve very high-frequency amplification in SGWB and, therefore, can only be accessible with the advent of future high-frequency SGWB detection techniques~\cite{Aggarwal:2020olq,Goryachev:2021zzn, Domcke:2022rgu,Bringmann:2023gba, Kondo:2024fvd, Valero:2024ncz,Carney:2024zzk,Domcke:2024mfu}. Indirect probes of ultra-low mass PBHs, such as DM or dark radiation from Hawking evaporation~\cite{Morrison:2018xla, Gondolo:2020uqv, Bernal:2020bjf, Green:1999yh, Khlopov:2004tn, Dai:2009hx,Allahverdi:2017sks, Lennon:2017tqq, Masina:2020xhk, Baldes:2020nuv,Hooper:2019gtx,Bhaumik:2022zdd,Cheek:2021odj,Cheek:2021cfe,Bernal:2022oha,Borah:2022iym, Unal:2023yxt,Lunardini:2019zob,Keith:2020jww,Cai:2020kfq,Masina:2021zpu,Arbey:2021ysg,Baker:2021btk,Calza:2021czr,Auffinger:2022khh,Thoss:2024hsr}, can also help distinguish memory burden effects from non-standard reheating history. We leave the detailed studies involving these directions for future works.

\section*{Acknowledgments}

NB wishes to thank his father Nikhil Kumar Bhaumik for his selfless support that never ceased during this work or any of NB's earlier works. The work of NB was supported by the International Center for Theoretical Physics Asia Pacific, the University of Chinese Academy of Sciences, and the National Science Foundation of China (NSFC) under Grant No. 12147103. MRH wishes to acknowledge support from the Science and Engineering Research Board (SERB), Government of India (GoI), for the SERB National Post-Doctoral fellowship, File Number: PDF/2022/002988. RKJ acknowledges financial support from the IISc Research Awards 2024, SERB, Department of Science and Technology, GoI through the MATRICS grant~MTR/2022/000821 and the Indo-French Centre for the Promotion of Advanced 
Research (CEFIPRA) for support of the proposal 6704-4 under the Collaborative Scientific 
Research Programme. 
The work of ML was supported by the Polish National Agency for Academic Exchange within the Polish Returns Programme under agreement PPN/PPO/2020/1/00013/U/00001 and the Polish National Science Center grant 2018/31/D/ST2/02048.
%%%%%%%%%%%%%%%%%%%%%%%%%%%%%%%%%%%%%%%%%%%%%%%%%%%%%%%%

\bibliographystyle{JHEP}
\bibliography{ref}
\end{document}